\documentclass[10pt, conference, compsocconf]{IEEEtran}

\usepackage{graphicx}
\usepackage[caption=false]{subfig}

\let\labelindent\relax

\usepackage{enumitem}

\usepackage{amsmath}
\usepackage{algorithm}
\usepackage{algorithmicx}
\usepackage[noend]{algpseudocode}

\usepackage{amsthm}
\usepackage{epsfig}
\def\IEEEbibitemsep{0pt plus .5pt}
\theoremstyle{definition}
\newtheorem{definition}{Definition}
\begin{document}
\bstctlcite{IEEEexample:BSTcontrol}
% \title{Composing Intermittent Energy Services%%\vspace{-3.5em}}

\title{Fluid Composition of Intermittent IoT Energy Services \vspace{-1.5 em}
}

%%---------------------AUTHORS---------------------------%%

\author{\IEEEauthorblockN{Abdallah Lakhdari}
\IEEEauthorblockA{School of Computer Science\\
The University of Sydney\\
Sydney, Australia\\
abdallah.lakhdari@sydney.edu.au}
\and
\IEEEauthorblockN{Athman Bouguettaya}
\IEEEauthorblockA{School of Computer Science\\
The University of Sydney\\
Sydney, Australia\\
athman.bouguettaya@sydney.edu.au}
}
\vspace{-1.5 em}

\maketitle
%%---------------------BODY---------------------------%%

% \begin{abstract}
% We propose a novel \textit{fluid} composition approach to crowdsource energy services in IoT environments. The proposed approach selects an optimal set of \textit{dynamic} services that fulfills the consumer's energy requirements \textit{wirelessly}. A new \textit{intermittent} service model is proposed to represent the dynamic crowdsourced energy services. We propose a new \textit{spatio-temporal heuristic} using the temporal knapsack algorithm for the fluid composition of intermittent services. Experimental results demonstrate the effectiveness and efficiency of the proposed approach.

%\end{abstract}

\begin{abstract}
We propose a novel \textit{fluid} composition approach of \textit{wireless} energy services in a crowdsourced IoT environment. The proposed approach selects an optimal set of \textit{dynamic} energy services according to the consumer's requirements. We leverage the {\em mobility patterns} of the crowd in confined areas to capture the intermittent behavior of IoT energy services. We model the IoT energy services based on their mobility patterns to propose a knapsack-based heuristic for the fluid composition. Experimental results demonstrate the efficiency of the proposed approach.
% A new \textit{intermittent} service model is proposed to represent the dynamic behavior of crowdsourced energy services. 

% We include the mobility patterns into energy service models to propose a new heuristic-based fluid composition.
% We propose a new  optimization approach 
% % technicality ....
\end{abstract}


% \fi
\IEEEpeerreviewmaketitle

\vspace{-0.2 cm}
\section{Introduction}
%%%%%%%%%%%%%%%%%%%%%%%%%%%%%%%%%%%%%%%%%%%%%%%%%%%%%%%%%%%%%%%%%%%%%%%%%%%%%%%%%%%%%%%%%%%%%%%%%%%%%%%%%%%%%%%%%%%%%%%%%%%%%%%%%%%%%%%%

The concept of the \textit{Internet of Things} (IoT) has emerged as a result of the pervasive presence of wireless sensors and embedded systems \cite{perera2014survey}. Physical \textit{things} are being connected to the Internet and augmented with sensors, computing resources, and network connectivity. These capabilities may be abstracted as \textit{IoT services} with functional and non functional properties (i.e., \textit{Quality of Service (QoS)}) \cite{bouguettaya2017service}. For example, sensing health-related information by a smartwatch is an IoT service where the functionality is sensing bio-signals from the body. The accuracy of signals and their freshness represent QoS attributes \cite{perera2014survey}.
%The service paradigm abstracts the IoT devices' capabilities 
% IoT devices' 
%service paradigm to define IoT services
%Examples
%%%%%%%%%%%%%%%%%%%%%%%%%%%%%%%%%%%%%%%%%%%%%%%%%%%%%%%%%%%%%%%%%%%%%%%%%%%%%%%%%%%%%%%%%%%%%%%%%%%%%%%%%%%%%%%%%%%%%%%%%%%%%%%%%%%%%%%%
% For example, a smartphone may provide its computing power as an IoT service \cite{ahabak2015femto}. 

% For example, sensing health related information by a smartwatch is an IoT service where the functionality is sensing the different bio-signals from the body. The accuracy of signals and their freshness  represent QoS attributes. 

%Any physical \textit{Thing} is being connected to the Internet. It is expected that by 2020, more than 50 billion things will be transformed into IoT devices ~\cite{perera2014survey}. These devices are augmented with sensors, computing resources, and network connectivity \cite{cabrera2018services}. 

%IoT services enable various applications in different domains, including smart cities, smart homes, and healthcare. 

%The \textit{service-oriented IoT} is a promising direction to develop an effective IoT ecosystem  \cite{cabrera2018services}.

%%%%%%%%%%%%%%%%%%%%%%%%%%%%%%%%%%%%%%%%%%%%%%%%%%%%%%%%%%%%%%%%%%%%%%%%%%%%%%%%%%%%%%%%%%%%%%%%%%%%%%%%%%%%%%%%%%%%%%%%%%%%%%%%%%%%%%%%%%%%

\textit{Crowdsourcing} leverages IoT services to create a self-sustained ecosystem. Crowdsourcing IoT services refers to sharing services among nearby IoT devices \cite{ren2015exploiting}. Examples of crowdsourced IoT services include computing offloading~\cite{ahabak2015femto}, environment monitoring \cite{zhang2015incentives}, and {\em wireless energy sharing} \cite{dhungana2020peer}. A smartphone with low battery power may for instance elect to receive energy from nearby {\em wearables using Wifi}. \textit{The focus in this paper is  on wireless energy sharing.}

The concept of {\em wireless energy sharing} (i.e., crowd charging) has been introduced recently to provide IoT users with {\em ubiquitous} power access through crowdsourcing \cite{bulut2018crowdcharging}\cite{dhungana2020peer}. Energy sharing provides a {\em green and convenient} alternative to charge IoT devices. For example, a smartwatch  can be recharged using the harvested energy of nearby smart shoes. The IoT devices can harvest energy from different sources, i.e., the kinetic movement of IoT users or their body heat \cite{gorlatova2014movers}. Several wireless energy transfer technologies allow the delivery up to 3 Watts power within 5-meter distances to multiple receivers e.g., \textit{Energous}\footnote{https://www.energous.com/} and \textit{Wi-Charge}\footnote{https://wi-charge.com/}.

% For example, a smart shoe may harvest energy from the physical activity of its user~\cite{choi2017wearable}. The harvested energy is able to recharge nearby IoT devices \textit{wirelessly}.

% It enables energy sharing between mobile IoT devices \textit{seamlessly} without the need for stationary power sources \cite{liu2016charging}. 

% Sharing energy wirelessly provides more \textit{convenience} to IoT users as they do not need to carry power banks.

% ~\cite{he2013energy,na2018energy}. 

% Several IoT devices manufacturers have already adopted the wireless charging technology \cite{lu2016wireless}. For example, the inductive coupling for wireless energy transfer between two smaptphones only allows a transmission  within millimeters or centimeters \cite{cook2017wireless}. Recently, a significant research is striving to support a Watt-level energy transmission over meter-level distance between IoT devices safely \cite{liu2016charging,zhang2017adaptive,fang2019fair}. Two companies, \textit{Energous} \cite{energous} and \textit{Wi-Charge} \cite{wicharge} have already produced their wireless charging product prototypes which can deliver up to 3 Watts power over 5 meters to multiple receivers. %This provides a perpetual energy supply for mobile devices. We propose to leverage the wireless  energy transfer to crowdsource wireless energy in an IoT environment \cite{na2018energy,lu2016wireless}. 

\textit{Crowdsourcing energy as a service} is the process of delivering wireless energy from IoT service {\em providers} to IoT users (i.e., {\em consumers}). Energy providers participate in the crowdsourced energy ecosystem for the following reasons. They might share energy altruistically to contribute to a green IoT environment. They can also be egotistic since energy is a vital resource for IoT devices. Therefore, providers would not be interested in sharing their energy unless they receive a satisfying incentive to compensate for their resource consumption. There is a body of research that considers incentives for crowdsourced IoT services \cite{zhang2015incentives}\cite{abusafia2020incentive}. In this paper, we focus on composing crowdsourced energy services. We assume that the providers are already incentivized by existing incentive models \cite{abusafia2020incentive}.

% \textcolor{blue}{Energy providers may participate in the crowdsourced energy ecosystem for a range of reasons. For example, they may elect to {\em altruistically} share energy to contribute to a green IoT environment. In contrast, the same green energy sharing outcome may be realized and motivated by {\em egotistic} purposes through a set of {\em incentives} \cite{zhang2015incentives}. An orthogonal aspect is the ability to provide a trusted framework for the crowdsourced IoT services among {\em a-priori} uknown {\em things}. New privacy-preserving trust models have been developed for crowdsourced IoT environments \cite{bahutair2019adaptive}. These two aspects are outside the scope of this paper. Our primary focus in this work is on composing Crowdsourced Energy Services (CES).}

%%%%TRUST and Privacy
{\em The composition} of IoT energy services is expected to play an   important role in the crowdsourced IoT environment \cite{Previouswork11}. A single energy service may not satisfy the consumer's requirement. The preferred solution is {\em to select an optimal set of services according to the requirement of the consumer} \cite{lakhdari2020composing}. Service composition has been extensively researched \cite{lemos2016web}. IoT services exhibit {\em flexible} and {\em dynamic} features. Conventional composition techniques may not be a good fit for energy services for the following \textit{distinct} characteristics:
\begin{itemize}[itemsep=0ex, leftmargin=*]
\item \textbf{Flexibility.} Energy service {\em consumers} do not have any lock-in contracts like traditional services. Assuming crowdsourced energy services are represented by time intervals, consumers may invoke services for the full-time interval or only partially according to their preferences.

\item \textbf{Intermittent behavior.} %Energy service {\em providers} do not have any Service Level Agreement (SLA) like cloud computing services \cite{ren2015exploiting}.
Energy providers may exhibit an {\em intermittent} behavior in a confined area at their advertised time. The energy providers do not necessarily commit to their initial advertisement. Energy services do not have any Service Level Agreement (SLA) \cite{lakhdari2020composing}. They may move freely inside the confined area. For example, an IoT user in a coffee shop may go to the counter to place an order and comeback \cite{waxman2006coffee}.
  
\item \textbf{Best-effort composition.} There is no failure if the composed services do not fulfill the exact requested energy amount. Any obtained amount is usable by the consumer. Unlike the binary traditional service composition \cite{medjahed2003composing}\cite{lemos2016web}, {\em the best-effort composition provides the best possible combination of services in terms of QoS properties and the consumer constraints}.

\end{itemize}

We introduce the concept of {\em fluid} composition considering the above characteristics of the crowdsourced IoT environment. The novelty of our proposed solution is that intermittent energy services are composed without any lock-in contracts or SLA. The fluid composition permits the partial consumption and re-invocation of intermittent services to collect the maximum amount of energy with the best QoS properties. If one of the component services {\em disconnects temporarily and comeback (i.e., intermittent behavior)}, the composition framework must decide between \textit{waiting} for the intermittent service or \textit{switching} to another service.

% We propose a {\em fluid} composition framework for crowdsourced energy services. 

% Let us assume a user requires a certain amount of energy to \textit{recharge} their devices, e.g., smartphone or smartwatch \textit{wirelessly} in a confined area such as food courts or coffee shops. Several crowdsourced energy providers may be available within the requesting user’s vicinity. The vicinity is defined by a predefined range which allows the wireless power transfer.  Energy providers may be observed by a set of QoS properties including distance, the provided energy, and the availability of the service.

% The fluid composition permits the partial consumption and re-invocation of intermittent services. If one of the selected component services disconnects temporarily and comeback (intermittent behavior), the composition framework must decide between \textit{waiting} for the intermittent service or \textit{switching} to another service.

% In this paper, we propose an \emph{offline} fluid composition approach to define the optimal selection of services ahead of time after estimating the mobility patterns and the cost of all disconnections and switches between available services. The main contributions of this paper are:

To the best of our knowledge, existing energy service composition approaches do not consider the mobility and the intermittent behavior of the crowdsourced energy service providers \cite{Previouswork11}. In this paper, we propose a novel model for \textit{mobile and intermittent crowdsourced energy services} which is a major extension of the deterministic energy service model in \cite{Previouswork11}. The indoor mobility patterns are predictable when people regularly visit confined areas such as restaurants and coffee shops \cite{waxman2006coffee}\cite{yang2015mobility}. We leverage these mobility patterns to propose a fluid composition approach. The proposed approach  finds an optimal selection of services ahead of time after estimating the mobility patterns and the cost of all disconnections and switches between available services. The main contributions are:   
\begin{itemize}[itemsep=0ex, leftmargin=*]
    \item A novel service model to represent the intermittent behavior of crowdsourced IoT energy services.
    \item A heuristic-based fluid composition algorithm to select an optimal set of intermittent IoT energy services.
%\item \textcolor{blue}{An experimental evaluation of the fluid composition framework based on scenarios close to reality.}
  
\end{itemize}
% The remainder of the paper is organized as follow: Section 2 presents the extended crowdsourced energy service model, QoS attributes, the energy query, and the composition problem. Section 3 details the proposed fluid composition algorithm. In Section 4, we discuss the experimental results. Some related works are presented in Section 5. Section 6 concludes the paper and highlights some future work. 
\vspace{-0.2 cm}
\section{Motivating scenario}
\vspace{-0.12 cm}
\begin{figure}
\centering
\includegraphics[width=0.32\textwidth]{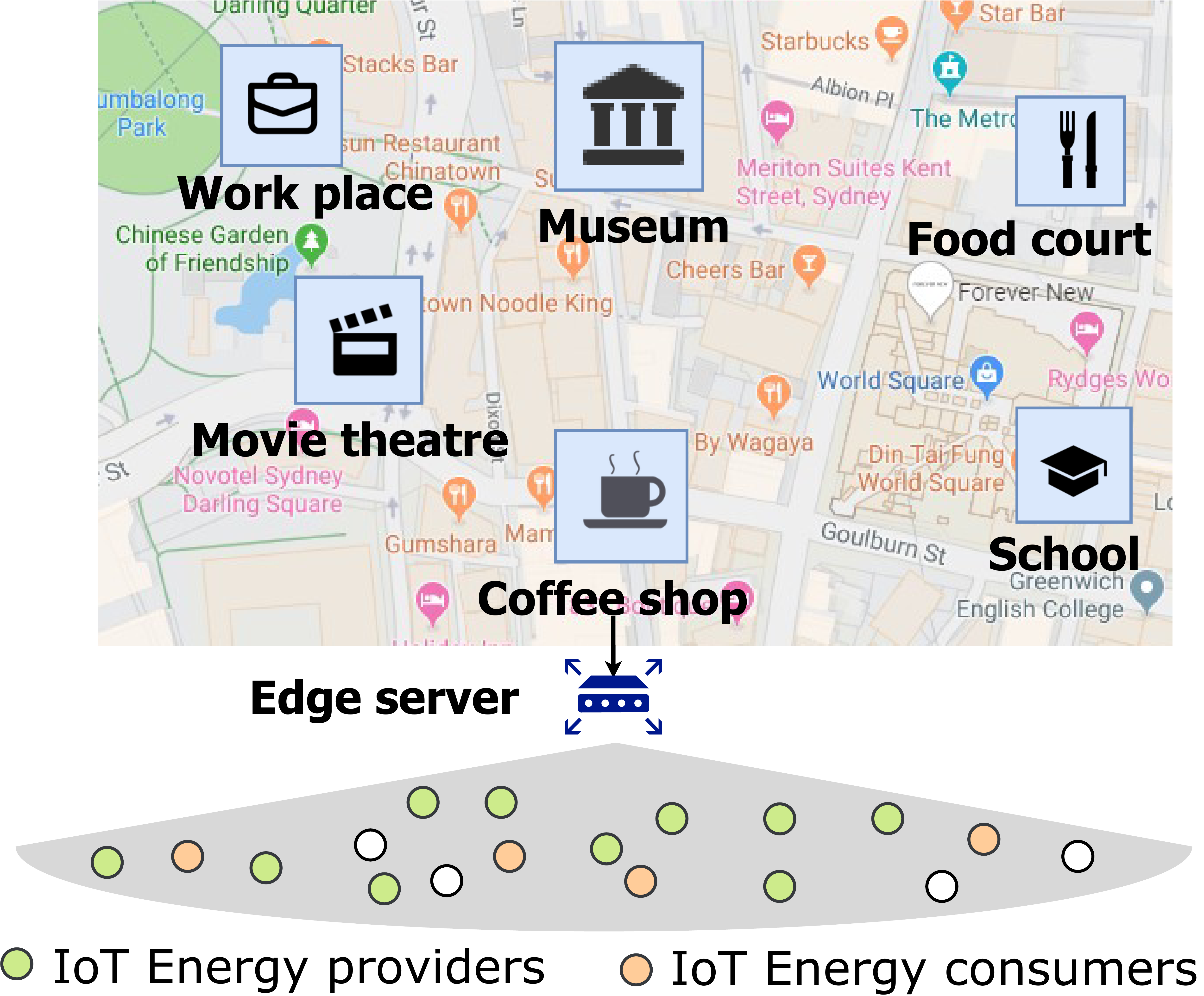}
\vspace{-0.1cm}
\caption{\small IoT users in a confined area in a smart city}
\vspace{-0.4cm}
\label{fig:deployments}
\end{figure}

% \begin{figure*}[t!]
%     \centering
%     \subfloat[]{\includegraphics[width=.4\textwidth]{Figures/motivationWISE2019aftercomments.pdf}}
%     \hfill
%     \subfloat[]{\includegraphics[width=.4\textwidth]{Figures/intermittentservices66666.pdf}}
%     %  %%%%\vspace{-0.35 cm}
%     % %%%\vspace{-0.1cm}
%     \caption{ Availability of intermittent services in the crowdsourced IoT environment}
%     % %%%\vspace{-0.4cm}
%     \label{fig:confidence}
% % %%%%\vspace{-0.6 cm}
% \end{figure*}
% \textcolor{red}{People may gather in confined areas e.g., coffee shop, restaurant, work space, theatre, etc. (see Figure \ref{fig:deployment} (a)). We assume that people have a set of IoT devices ({\em wearables}) which may harvest energy \cite{choi2017wearable}. They may also share their spare energy wirelessly with nearby IoT devices \cite{bulut2018crowdcharging}. In a confined area, we assume that there are a set of energy service providers and consumers (see Figure \ref{fig:deployment} (b)).}

% \textcolor{blue}{
We describe a scenario where a number of people gather in different places, i.e., {\em confined} areas (e.g., coffee shops, restaurants, movie theatres) within the downtown of any major city (see Figure \ref{fig:deployments}). 
We assume that city dwellers will use their {\em wearables} to harvest energy \cite{khalifa2017harke}. Note that existing wearable technology allows for the harvesting of several microwatts to a few watts \cite{gorlatova2014movers} \cite{khalifa2017harke}. Users wearing a harvester on each leg can generate enough power to charge up to four smartphones by walking for one hour at a comfortable pace e.g., PowerWalk\footnote{https://www.bionic-power.com/}. This provides an opportunity for IoT devices to share spare energy wirelessly with nearby devices. The distance between IoT devices exchanging energy may reach five meters to ensure a successful wireless transmission \cite{dhungana2020peer}. The IoT devices and wearables are assumed to be equipped with wireless energy transmitters and receivers using products such as Energous and Wi-Charge.
% } 

\begin{figure}
\centering
\includegraphics[width=0.42\textwidth]{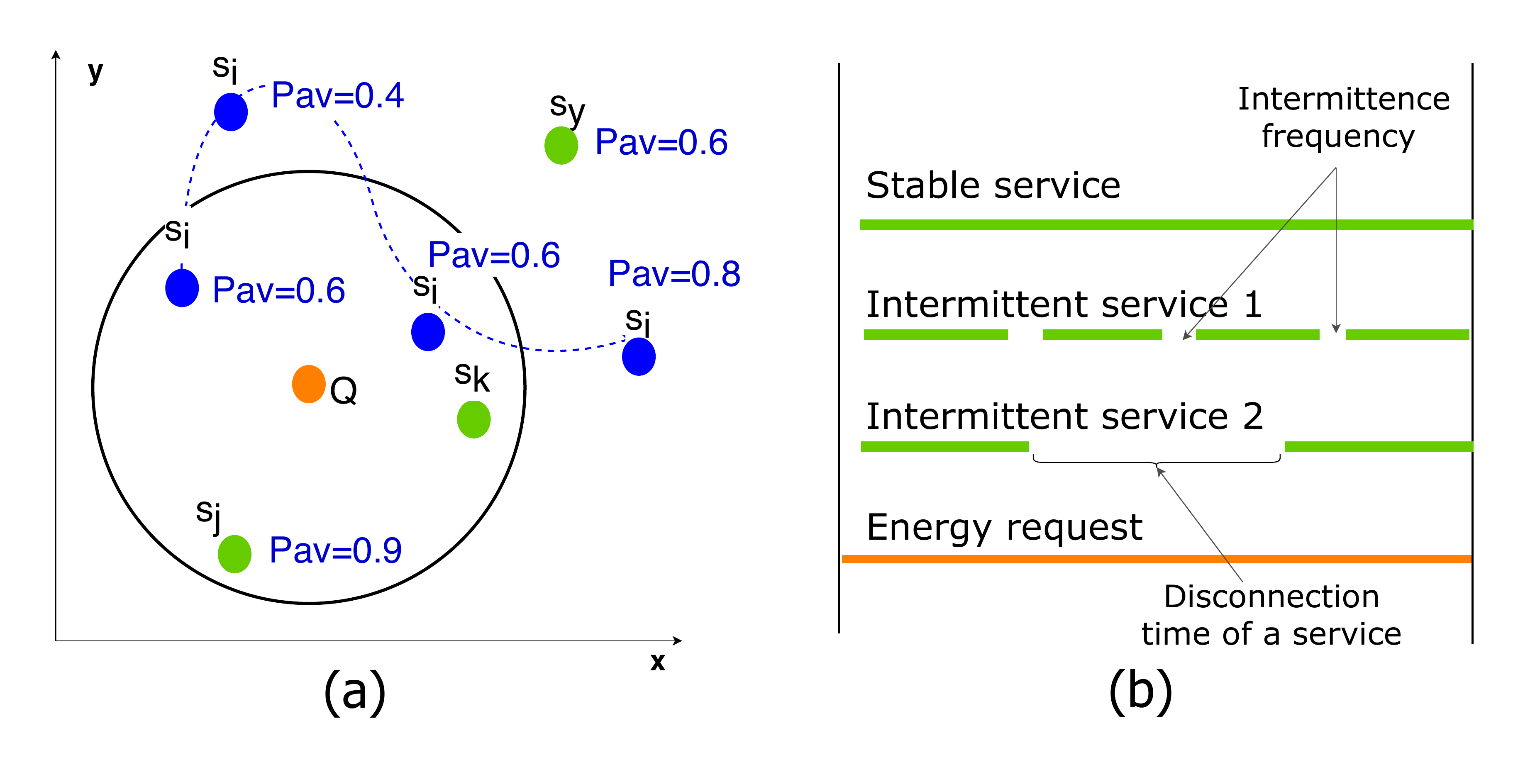}
\vspace{-0.4cm}
\caption{\small Availability of intermittent services in the crowdsourced IoT environment (a) Spatial patterns of IoT energy services (a) Temporal paterns of IoT energy services}
\vspace{-0.4cm}
\label{fig:confidence}
\end{figure}

% We also assume that all IoT devices utilize an accurate geolocalization system such as ultra wide band based indoor positioning system\footnote{https://www.pozyx.io/}.} 
%Let us now consider a scenario of 
We will use the following scenario: An IoT user in a food court needs to \textit{recharge} their smartphone to be able to run some important applications, e.g., make a call or use email. The IoT user launches the following request: \textit{User $x$ requires an amount of energy $E$  in the location $L$ during the period $[st~,~et]$}. All energy requests are processed by a centralized edge-based IoT coordinator, i.e., a router in a confined area (see Fig. \ref{fig:deployments}). Multiple IoT energy service providers are advertising their energy services in the food court at the same time. The advertisement presents information about the provided service, e.g., the service location, start time and end time, the provided energy amount. Note that, the wireless energy transfer can be performed only in a predefined range between the providing and the consuming IoT devices \cite{dhungana2020peer}. A neighboring energy service may \textit{drop out temporarily} at any moment due to the mobility of device owners, e.g., an IoT user may go out to have a phone call and come back (see Figure \ref{fig:confidence} (a)). Therefore, the {\em intermittent mobility} of IoT devices affects the validity of any composition of  available IoT energy services as the delivery of the expected amount of energy may not be guaranteed. The component services may not provide the exact advertised energy amount because of their disconnections.% after moving away from the predefined energy wireless delivery range. 

\textit{Typically, people develop certain \textit{routines} in confined areas} \cite{waxman2006coffee}. They tend to define a set of preferences inside a specific confined place. For example, a user may select a preferred seat in the food court. They would sit and chat for a while. The user then would go to order food, they would wait in the line, order, and come back to their seat. The user goes back to receive the order. Finally, they leave the food court after finishing the food. {\em The place preferences and habits build a mobility pattern for IoT users in confined areas}. Capturing the mobility of energy providers permits us to estimate the real availability of crowdsourced energy services. Several prediction models are proposed to capture the indoor mobility patterns of the crowds \cite{yang2015mobility}. Estimating the availability of intermittent services within the range of the consumer allows the selection of a high-accurate composition of crowdsourced energy services in a dynamic environment. The effective availability of intermittent services is affected by the frequency and the length of the temporary disconnections (see Figure \ref{fig:confidence} (b). The \textit{reactive composition} which considers all the disconnections and switches at every disconnected service to a new one provides a high-accurate solution. However, the reactive composition \textit{lacks runtime efficiency} \cite{lemos2016web} in the highly dynamic crowdsourced environment. The reactive composition  is also costly in terms of time and energy for every new connection establishment \cite{lakhdari2020composing}\cite{na2018energy}. An edge-based IoT coordinator requires a lightweight composition framework to deal with dynamic IoT energy services. We formulate our composition problem as \textit {"finding the optimal composition that solves the trade-off between effectiveness and runtime efficiency."}
% An edge-based energy crowdsourcing framework requires a lightweight composition algorithm to deal with dynamic IoT energy services. 

\vspace{-0.22 cm}
\section{Intermittent energy service model}\label{sysmdlpb}
\vspace{-0.12 cm}
The intermittent crowdsourced energy services are provided by peoples' wearables \cite{gorlatova2014movers}. The mobility pattern of IoT users in a smart city allows the estimation of energy services' availability \cite{gonzalez2008understanding}. However, the available services may deliver energy intermittently due to their movement within the confined area (i.e., micro-mobility).  The micro-mobility of IoT energy providers may cause disconnections of the wireless energy delivery. Each individual may concur a unique experience in a particular place. This experience may create an attachment of people to that particular place reflected in a set of habits \cite{waxman2006coffee}. These habits may define the {\em mobility pattern} of individuals in confined areas \cite{yang2015mobility}. We leverage mobility patterns to define the intermittent energy service model.
% We define the \textit{micro}-mobility by the movement of people in confined areas such as coffee shops, restaurants, and theatres \cite{yang2015mobility}. The availability of a service in a coffee shop depends on the mobility of the IoT device owners.  

% The available services may deliver energy intermittently due to  

% two spatio-temporal patterns of crowdsourced energy services based on the mobility of the crowd in two different scales, i.e., a) \textit{macro}-mobility, and b) \textit{micro}-mobility of the crowd. The macro-mobility indicates the movement of people in a large geographic area e.g., city and suburb.  
%\vspace{-0.07 cm}
% \theoremstyle{definition}
\begin{definition}{An intermittent crowdsourced IoT energy service} \emph{ $ CES $ is represented as a tuple $< Eid, Eownerid, F , Q, A, In>$ where $Eid$ is a unique service ID, $Eownerid$ is a unique ID for the owner of the IoT device, $F$ is the set of $CES$ functionalities offered by an IoT device. $Q$ is a tuple of $<q_1, q_2, ..., q_n>$ \cite{Previouswork11}. $A$ and $In$ are the spatio-temporal patterns which capture the mobility of energy services. Each $q_i$ denotes a QoS property of $CES$. $A$ is a temporal function for the estimation of the availability of $CES$ inside the confined area at the advertised time. $In$ is time series to represent the intermittent behavior of $CES$.}
\end{definition}

% %%%%\vspace{-.15cm}
% \paragraph{Definition 2.}
% %%%\vspace{-0.25 cm}
%\vspace{-0.07 cm}
\begin{definition}{Crowdsourced Energy Service Consumer request.}
\emph{An energy service request is defined as a tuple $Q=< t, l, RE, CI, du>$, where $ t$ refers to the timestamp when the request is launched. $ l $ refers to the location of the consumer. We assume that the consumer is stationary at their location $l$  after launching their energy request. $RE$ represents the required amount of energy. $CI$ is the maximum intensity of the wireless current that a consuming IoT device can receive. $du$ refers to the charging period \cite{Previouswork11}}.
\end{definition}
% \textcolor{blue}{In the rest of the paper, $"."$ means {\em "element of"} e.g., $Q.RE$ is the required energy $RE$ by the query $Q$.}

\begin{definition}{Crowdsourced Energy Service Quality Attributes.}
\emph {Quality parameters i.e., QoS allow users to distinguish among crowdsourced IoT energy services.
QoS parameters are defined as a tuple $< l, r,  st, et, DEC, I, Tsr, Rel_i>$: $l$ is the location of the consumer. $r$ is the range between the providing and consuming IoT devices which allows a successful wireless energy delivery. $st$ and $et$ represent the start time and end time of a crowdsourced energy service respectively. $DEC$ is the deliverable energy capacity. $I$ is the intensity of the wirelessly transferred current. $Tsr$ represents the transmission success rate. $Rel_i$ is the reliability QoS \cite{Previouswork11}}. 
%incentive and tust assumptions
\end{definition}
\vspace{-0.22cm}
The spatio-temporal features of the IoT energy services (i.e, $l, st $ and $et$) are defined based on the pattern of time spent in regularly visited places e.g., coffee shops or food courts using their daily activity model \cite{gonzalez2008understanding}. $DEC$ and $Rel_i$ are estimated by the energy usage model of the IoT device \cite{bulut2018crowdcharging}.  $I$ and $ Tsr$ are defined based on the specifications of the IoT devices providing services. The intermittent mobility patterns of IoT energy services are defined as:
\begin{itemize}[itemsep=0ex, leftmargin=*]
    \item \textit{\textbf{Availability}}: The availability of an intermittent  service is the probability distribution of service provider's location inside 
    % the range of energy transmission to the consumer 
    the confined area $C$ during the advertised time interval.% \paragraph{Definition 3.} \textit{Availability of crowdsourced energy services} the availability of a service is the probability of presence of an energy service in a confined area during the advertised time interval.
% The availability of crowdsourced energy services depends mainly on the mobility of the IoT device owners. The services are not necessarily available at the exact advertised space and time. 
We denote the  availability $A$ of a service $i$ within a confined area $C$ as a tuple $A_i(C, t_{i}, loc_{i}, \theta_{i})$. %, n=0,1,2,~..,k$. 
Here, $t_i=\{t_{i0},t_{i1},t_{i2},~..,t_{in}\}$ is the set of timestamps between the start $st_i$ and the end time $et_i$ of service $i$. $loc_i=\{loc_{i0},loc_{i1},loc_{i2},~..,loc_{in}\}$ is the set of locations of the service $i$ within the confined area $C$.  Each $\theta_{ik} \in \theta_i=\{\theta_{i0},\theta_{i1},\theta_{i2},~..,\theta_{in}\}$ is the probability that the service $i$ is in the location $loc_{ik}$ at the timestamp $t_ik$, $\theta_i(l_i=loc_{i})= f(H_i,C)$.
The probability distribution function $f$ can be obtained by statistical methods applied on the historical records $H_i$ of service $i$ in the confined area $C$ \cite{deng2016constraints}.
% The probability distribution function can be declared by
% the providers or obtained by some statistical learning methods. Service providers may report their estimated stay time to
% the service repository wishing their services to be invoked
% more often and create more benefit. There exists similar incentive mechanisms in mobile ad hoc networks to obtain cooperative routing and forwarding [9]. Alternatively, the stay
% time of a service provider can also be obtained without explicitly requesting such information from the provider. For
% example, in the example of Section II, the arrival time of the
% subway can be predicted to a time window and the probability
% of each time point can be computed according to the historical records. 

%service many times defines pattern

%given the advertisement of $CES$ in a confined area $C$ during the time interval $[st_i,et_i]$.

\item \textit{\textbf{Intermittent provision}}: The intermittent provision $In$ of  an IoT energy service $i$ to an energy request $Q$ is modeled as a tuple $In_i(Q, A_i, Pr_i)$  where $A_i$ is the availability distribution for all timestamps $t_i=\{t_{i0},t_{i1},t_{i2},~..,t_{in}\}$ between the start $st_i$ and the end time $et_i$ of service $i$. $Pr_i$ is the corresponding energy provision distribution for all timestamps based on the location of the service. %The intermittent time series $In_i(Q, A_i, Pr_i)$ 
The provision distribution $Pr_i$ presents the wireless energy provision status at each timestamp $t_{ik} \in t_i=\{t_{i0},t_{i1},t_{i2},~..,t_{in}\} $ as follows:   
%explain formally by range and distance boolean
% \[
%     Pr_{ik}= 
% \begin{cases}
%     1,&  D_k(Q.l,l_i)\leq r_i\\
%     0,& otherwise
% \end{cases}
% \]
$$Pr_{ik}=1 ~if~  D_k(Q.l,l_i)\leq r_i;~~0 ~ otherwise $$
Where $D_k(Q.l,l_i)$ is the distance at timestamp $t_k$ between the location of the energy consumer $Q.l$ and the location $l_i$ of the service $i$. $r_i$ is the wireless transmission range of the device providing the service $i$.
% which allows a successful wireless energy delivery. 
%time-span when the IoT user is available and providing their energy service in the confined area. It is represented as a set of continuous time points. 
% $L_i$ is the set of the user's locations corresponding to all time points in $T_i$, assuming that users are moving inside the confined area. 

    % \item $Pret$ is the probability of returning to the initial distance to the consumer after temporary disconnection.
%\end{enumerate}

% %%%%\vspace{-.4cm}
 
\end{itemize}
\vspace{-0.2 cm}
\section{Composition of Crowdsourced IoT \\ energy services}
\vspace{-0.12cm}
%In this section, we develop a new fluid composition algorithm to compose intermittent crowdsourced energy services. We first define fluid composition as follows. 

\begin{definition}{Fluid composition.}
%\subsection{Crowdsourced energy service composition problem}
%%%%\vspace{-0.2 cm}
%Providers advertise their services without any SLA or lock-in contracts. Usually, people move freely inside confined areas, e.g., coffee shop and restaurant. This movement exhibits an intermittent service provision in such areas. 
%\paragraph{Definition 5} \textit{spatio-temporal crowdsourced energy service composition problem}
\emph{Given a set of intermittent crowdsourced IoT energy services; $S_{CES}=\{ CES_{1}, CES_{2}, \dots CES_{n} \} $ and a request $Q=< t, l, RE, CI, du>$ in a confined area $C$, the fluid composition framework relies only on the advertisement and the estimation of the mobility patterns to select and compose IoT energy services \textit{ahead of time}.
% is to find the optimal composition of nearby  energy services $ CES_{i} \in S_{CES}$ \textit{ahead of time} \textcolor{blue}{i.e., The composition framework relies only on the advertisement and the estimation of the mobility patterns to select and compose energy services}.
The optimal composition can transfer the required amount of energy $RE$ considering the intermittent behavior of nearby services.} 
\end{definition}
We transform the fluid composition problem into a temporal knapsack problem \cite{bartlett2005temporal} as follows:
%\vspace{-0.14cm}
 
Maximize $Composite.DEC = \Sigma CES_i.DEC$
%\vspace{-0.05cm} 

Where  $Composite$ is the composition of intermittent $CES$. $CES_i$ are the component services of $Composite$. We consider the following assumptions: 
\begin{itemize}[itemsep=0ex, leftmargin=*]
    % \item All the selected time intervals for component services $CES_i$ i.e.,  $[st_i,et_i]$ must be within the request duration $[t,t+du]$.

    \item The providers may move freely inside the confined area $C$. This movement leads to the intermittent wireless IoT energy delivery.
    \item Energy can be be provided wirelessly if and only if the distance between the providing and consuming devices is equal or less than a predefined range $r$.
    \item Two or more energy services can be composed \textit{simultaneously} if and only if the sum of their current transfer intensity $I_i$ is lower or equal the intensity of the request. $\Sigma I_i \leq CI$ \cite{na2018energy}.
    \item We assume that it is possible to invoke services partially. 
\end{itemize}

\begin{figure}
    \centering
    \includegraphics[width=.42\textwidth]{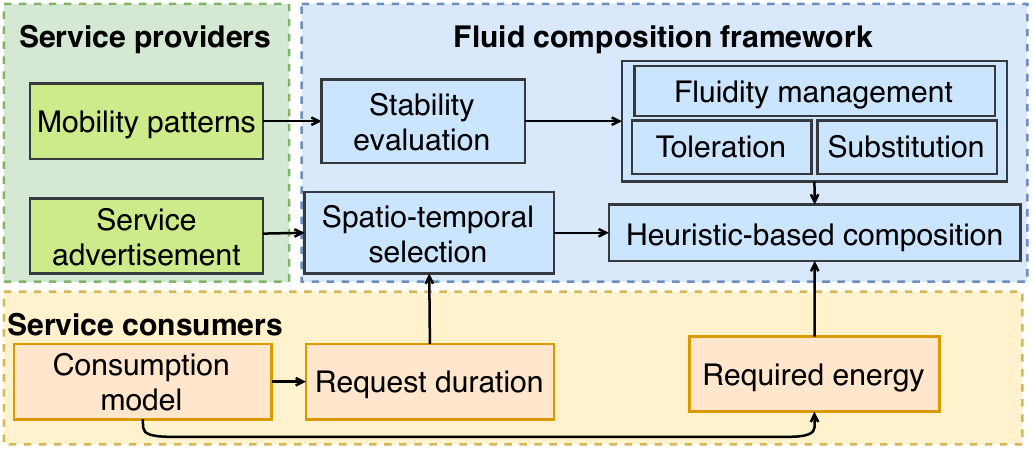}
    %\vspace{-0.1cm}
    \caption{ \small Fluid composition framework}
    \vspace{-0.4cm}
    \label{fig:frmwrk}
\end{figure}

The temporal knapsack algorithm composes crowdsourced IoT energy services based on their advertisement. However, Energy services typically exhibit an intermittent behavior. We leverage the spatio-temporal patterns captured by the proposed service model to present a novel framework for composing intermittent energy services (see figure \ref{fig:frmwrk}). The framework consists of three phases: (i) It starts by selecting the composable IoT energy services based on the request duration and the advertisement of energy providers. (ii) The framework then evaluates the intermittent behavior of candidate energy services based on their mobility patterns. (iii) The proposed heuristic composes an optimal set of candidate energy services \textit{ahead of time} as follows: {\em Tolerating} the stable but slightly intermittent candidate services. {\em Substituting the disconnected services by the most stable available services}.

%The proposed composition is an algorithm which selects the most stable composition \textit{ahead of time} based on the intermittent patterns of available services. In this paper, \textit{we assume that services have the same probability of being available inside the confined area}. Our future work will consider the macro-mobility and the probability distribution of the providers' availability. We design our composition technique based on the intermittent behavior of services. We propose \textit{evaluation metrics} of crowdsourced energy services according to their mobility patterns. We utilize these metrics to select the optimal composite service fulfilling the energy requirement of an IoT user.

\vspace{-0.22cm}
\subsection{Spatio-temporal selection of IoT energy services}
\vspace{-0.1cm}
% We define the spatio-temporal composition plan of crowdsourced energy services in a predefined energy request duration. 
Initially, we select and compose the IoT energy services according to their spatio-temporal features and the request duration \cite{Previouswork11}\cite{lakhdari2020composing}. The spatio-temporal selection algorithm consists of three steps:  
 
% we propose a modified temporal knapsack algorithm \cite{bartlett2005temporal} for the spatio-temporal selection  of services. The algorithm consists of three steps: 
%\vspace{-0.17cm}
% Next, we present heuristics to select the optimal candidate intermittent energy services for the fluid composition.

%%%\vspace{-0.4cm}
% \subsubsection{Spatio-temporal composition of  energy services.}
% We propose a modified temporal knapsack algorithm \cite{bartlett2005temporal} for the spatio-temporal composition of services. The algorithm consists of three steps: 
% \begin{figure*}[t!]
%     \centering
%     \subfloat[]{\includegraphics[width=.35\textwidth]{Figures/InitialChunks.pdf}}
%     %\hfill
%     \hspace{1.5 cm}
%     \subfloat[]{\includegraphics[width=.35\textwidth]{Figures/FinalChunks.pdf}}
%     %  %%%%\vspace{-0.35 cm}
%     %%%\vspace{-0.1cm}
%     \caption{ Temporal chunking of crowdsourcing energy services}
%     %%%\vspace{-0.4cm}
%     \label{fig:Chunking}
% % %%%%\vspace{-0.6 cm}
% \end{figure*}
\begin{enumerate}[itemsep=0ex, leftmargin=*]
    \item We select the composable services based on their advertisement. All available services within the request duration are considered {\em temporally composable} services. Selected services must be within the range of the energy consumer to be {\em spatially composable} and allow the wireless energy transfer \cite{lakhdari2020composing}. 

    \item We chunk the request duration into smaller temporal chunks based on the advertisement of composable service. We assume that energy services can be decomposed and consumed \textit{partially}. The energy consumer may switch to other energy services within the request duration. We define all the possible timestamps where the consumer may switch to another service. Each timestamp is either the start or the end time of available services \cite{Previouswork11}.
    
    % (see Figure \ref{fig:Chunking}(a)). 

    \item  We apply the 0/1 knapsack algorithm at each chunk by considering the wireless intensity compatibility between the consuming and providing IoT devices. We maximize the acquired energy at each chunk  by combining the partial services with respect to their wireless current intensity, i.e., the aggregated intensity of composed partial services at a chunk must be lower or equal the compatibility intensity $CI$ of the consumer \cite{lakhdari2020composing}.
\end{enumerate}

\vspace{-0.22cm}
\subsection{Evaluation of intermittent IoT energy services}\label{ftrs}
\vspace{-0.1cm}
% \begin{figure*}[t!]
%     \centering
%     \subfloat[]{\includegraphics[width=.35\textwidth]{Figures/InitialChunks.pdf}}
%     %\hfill
%     \hspace{1.5 cm}
%     \subfloat[]{\includegraphics[width=.35\textwidth]{Figures/FinalChunks.pdf}}
%     %  %%%%\vspace{-0.35 cm}
%     %%\vspace{-0.1cm}
%     \caption{ Temporal chunking of crowdsourcing energy services}
%     %%\vspace{-0.4cm}
%     \label{fig:Chunking}
% % %%%%\vspace{-0.6 cm}
% \end{figure*}

%%%%%%%%%%%%%%%%%%%%%%%%%%%%%%%%%%%%%%%%%%%%%%%%%%%%%%%%%%%%%%%%%%%%%%%%%%%%%%%%%%%%
% \begin{figure}
%     \centering
%     \includegraphics[width=.30\textwidth]{Figures/FinalChunks.pdf}
%     %\vspace{-0.1cm}
%     \caption{ Temporal chunking of energy services}
%     %\vspace{-0.4cm}
%     \label{fig:Chunking}
% \end{figure}

%%%%%%%%%%%%%%%%%%%%%%%%%%%%%%%%%%%%%%%%%%%%%%%%%%%%%%%%%%%%%%%%%%%%%%%%%%%%%%%%%%%%
%%%\vspace{-0.3cm}
% The service model captures the intermittent behavior of crowdsourced energy services. 
The service model utilizes the mobility pattern $A_i(C, t_{i}, loc_{i}, \theta_{i})$ to capture the disconnections whenever the energy service $i$ moves inside the confined area $C$. The intermittent provision pattern $In_i(Q, A_i, Pr_i)$ defines the disconnections between the service $i$ and the request $Q$. The frequency of disconnections reflects the stability of the wireless energy provision. Energy services are more stable when the disconnections are less frequent. We define a \textit{stability score} $STB_i(Q)$ for a service $i$ toward an energy request $Q$  by the frequency of disconnections as follows: 
%\begin{equation}
     %\small
     $$STB_i(Q)= 1-\frac{1}{|et_i-st_i|-\sum_{0}^{n}Pr_{ik}}, k\in t_{i}$$ 
%\end{equation}
 Where, $st_i$ and $et_i$ represent the start and end time of service $i$. $Pr_{ik}$ is the provision status of service $i$ to the energy request $Q$ at timestamp $k\in t_{i}$. $t_{i}$ represents all the time points for the time interval of service $i$. 
 
The length of a disconnection also affects the energy provision. An energy consumer sets their request for a period of time $Q.du$. Long service disconnections cause considerable loss according to the consumer temporal preferences. \textit{We also evaluate services based on their disconnection time}. $ADis_i$ is the disconnection time ratio. It represents the accumulated disconnection time relatively to the initial service availability time.
% Where, $f_{dis}(S)$ is the number of disconnections of service $S$ from the IoT device requesting energy $Q$. The length of a disconnection also affects the energy provision. An energy consumer sets their query for a certain period of time. Long service disconnections cause considerable loss according to the consumer temporal preferences. \textit{We also evaluate services based on their disconnection time}. $G_{dis}$ is the disconnection time ratio. It represents the disconnection time relatively to the initial service availability time.
%\vspace{0.4cm}
%\begin{equation}
$$ ADis_i(Q) = \frac{\sum_{m} dis_m}{|et_i-st_i|}$$
%\end{equation}
%|t_{ik}-t_{i(k+dis^m)}|
where $DIS_i=\{dis_1, dis_2, .., dis_n \}$ are the disconnections of service $i$ from the energy request $Q$. Any time interval $\tau[m, m+dis] \in [st_i,et_i]$ is considered as a disconnection if and only if:
$\tau \in DIS_i \iff Pr_im=0 $ $\land Pr_{i(k+dis)}$ $\land \forall j\in \tau, Pr_ij=0$

We utilize these two metrics to select the best candidate energy services in the fluid environment to provide the optimal spatio-temporal composition.% fulfilling the energy requirement of an IoT user in a predefined period of time.

\vspace{-0.2cm}
\subsection{Fluid composition of crowdsourced IoT energy services}
\vspace{-0.12cm}
\begin{algorithm}[t!]
\footnotesize
    \renewcommand{\algorithmicrequire}{\textbf{Input:}}
    \renewcommand{\algorithmicensure}{\textbf{Output:}}
    \caption{Heuristic-based fluid composition}
    
    \label{alg:STcompo}
    \begin{algorithmic}[1]
        \Require
        $ Q.l$, $Q.t$ , $Q.du$, $Q.CI$, $NearbyS$ 
        \Ensure $Composite$ component energy services  during $Q.du$
        %step 1
        \Statex \text{ // Defining substitute services}
        \For{$ S_i \in NearbyS $}
            \If{$STB_i(Q)\leq \mu $}
                \If{$Adis_i(Q) \geq D$}
                    \State $NearbyS$.remove($i$)
                \Else
                    \Statex \text{ // look for substitute services}
                    \State {$Subs_i= \emptyset$}
                    \For{All $ dis \in i $ and $|dis.st-dis.et|\geq G$}
                        \State {$Subs_i= Subs_i \cup $ Find services($dis.st, dis.et$)}
                        \Statex \text{ // merge substitutes with initial service $i$}
                        \State {$i= i \cup Subs_i$}
                     \EndFor
                \EndIf 
            \EndIf
        \EndFor
        
        % Define stable services services
        %step 2
        \Statex \text{ // Chunking the query based on the advertisement of $NearbyS$ services}
		\State $Chunk_{0}.st \gets Q.t $
		\For{$ int~t= Q.t~to~ Q.t+Q.du $}
		    \If {$(\forall~CES\in NearbyS~and~t =st_CES ~or~t = et_CES)$}
		        \State $ Chunk_{i}.et\gets t $
		    \EndIf
		    \Statex \text{ // create new chunk}
		    \If {$t \neq Q.t+Q.du  $}  
		        \State $ Chunk_{i+1}.st\gets t $
		        \State $ t \gets t+1 $
		    \EndIf
		\EndFor
        %step 3
        \Statex \text{// apply 0/1 knapsack optimization at each chunk}
        \For{   $C\in Chunk$}
        	\Statex \text{// $miniComposite$ is the local composition in a chunk}
		    \Statex \text{// $miniCES$ is the set of  partial service within a chunk}
		    \State $miniComposite \gets \emptyset$
		    \State \textbf{While}\text{($miniCES \neq \emptyset$)}
		        \State    \text{$Smax  \gets Max( ~miniCES )$} 
		        \State    $miniCES  \gets miniCES~-~\{Smax\}$ 
 		        \If        {($ Composable(minicomposite~,~ max) $ ) }
 		                \State    \text{$miniComposite \gets miniComposite \cup \{max\}$}
 		        \EndIf
 		    \State \textbf{End While}
 		    \State \text{$Composite \gets Composite \cup \{ miniComposite\}$}
        \EndFor
        \State \Return $ Composite $
    \end{algorithmic}
\end{algorithm}
The fluid composition composes the intermittent energy services based on their {\em stability} and {\em disconnection} metrics. \textit{We propose the following modifications of the spatio-temporal composition algorithm to incorporate intermittent energy services}: 

% The proposed temporal knapsack algorithm composes crowdsourced energy services based on their advertisement. Energy services typically exhibit an intermittent behavior. The proposed service model permits to capture the spatio-temporal patterns of intermittent services and identify their disconnections. 

\begin{figure}
    \centering
    \includegraphics[width=.4\textwidth]{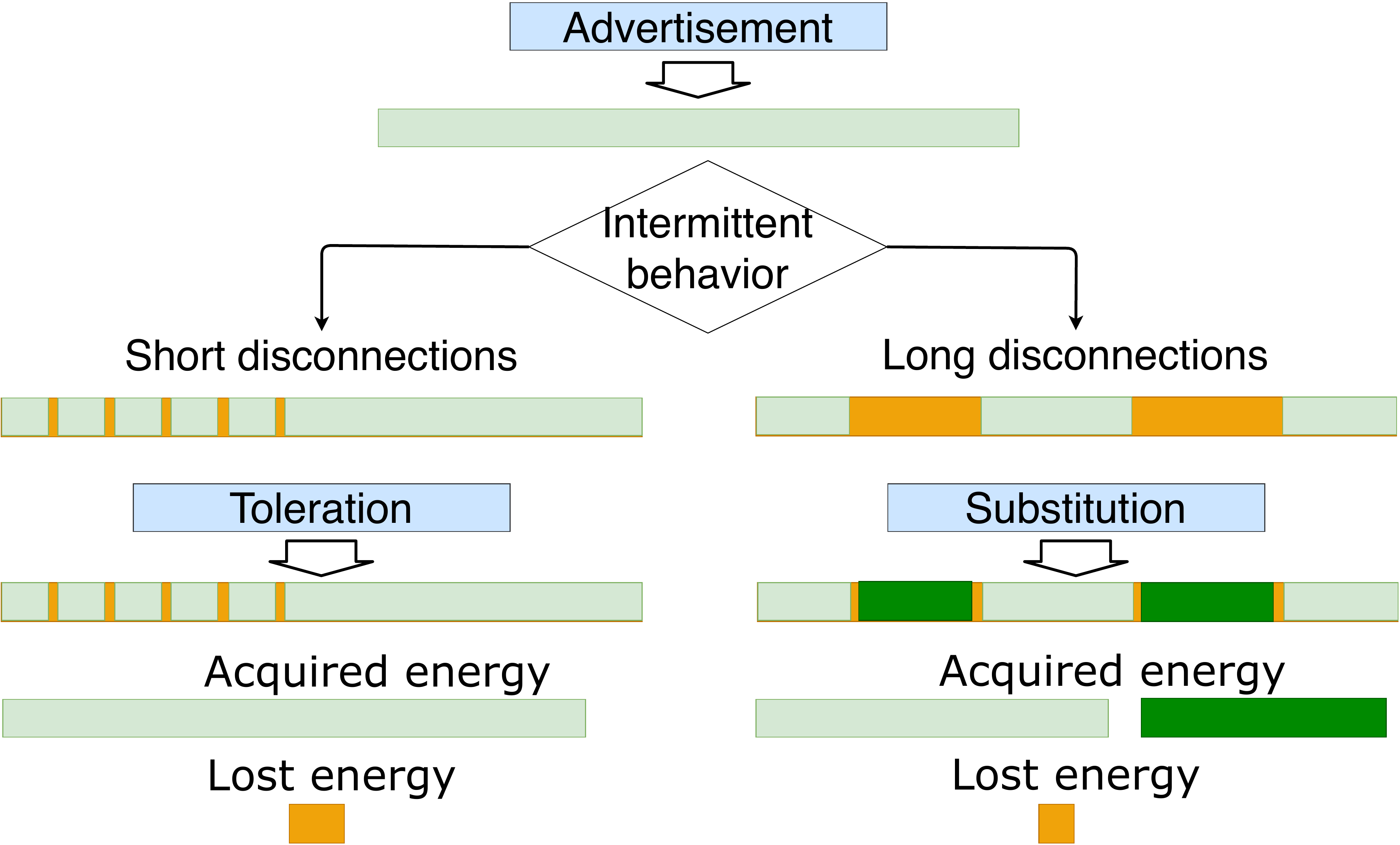}
    \vspace{-0.1cm}
    \caption{ \small Heuristics for the fluid composition}
    \vspace{-0.5cm}
    \label{fig:heuristic}
\end{figure}

%\vspace{-0.22cm}
\subsubsection{Brute-force spatio-temporal fluid composition} 
{\em The disconnected services may be considered as a set of independent services}. The temporal knapsack composition \cite{Previouswork11}\cite{deng2016constraints}  can be utilized on this new set of disconnected services.
%(see Figure \ref{fig:Chunking}). 
However, this composition technique is inefficient and very costly in a dynamic environment. A considerable amount of energy is lost because of the increasing number of switches between partial services. Every switching triggers a new connection establishment between the consuming and a providing device which requires energy \cite{na2018energy}\cite{lakhdari2020composing}. The computing time increases significantly by applying the 0/1 knapsack algorithm at every new chunk. 
% It is required to find an efficient composition technique for intermittent services.    
%%%\vspace{-0.4cm}
% \subsubsection{Lossy spatio-temporal fluid composition} 
% Intermittent crowdsourced energy services may be selected and composed according to their intermittence features, stability score and the accumulated disconnections length. The lossy composition algorithm removes all intermittent services below a predefined threshold from the set of composable services. The lossy composition selects only stable services. Similar to the Greedy Perimeter Stateless Routing GPSR algorithm \cite{karp2000gpsr}  in Wireless Networks which selects the closest node services, the lossy spatio-temporal composition considers only the most stable nearby services. This selection criteria limits the performance of the spatio-temporal composition in a dynamic environment where intermittent services are inherently mobile.

% \textcolor{red}{[Please write more, what is lossy, why you consider it, an example, how to connect intermittence features, stability score and the accumulated disconnections length]}
%\vspace{-0.22cm}
\subsubsection{Heuristic-based fluid composition}
% %%\vspace{-0.15cm}
\begin{table*}[t!]
\centering
\small
\begin{tabular}{|l|c|c||l|c|c|}
\hline
\multicolumn{3}{|c||}{\footnotesize{Crowdsourced energy service}}                                                                                                          & \multicolumn{3}{c|}{\footnotesize{Energy query}}                                                                                                                                                                                  \\ \hline
\multicolumn{1}{|c|}{\footnotesize{QoS}} & \footnotesize{Dataset}                                                              & \footnotesize{value}                                                       & \multicolumn{1}{c|}{\begin{tabular}[c]{@{}c@{}}\footnotesize{Query parameters}\end{tabular}} & \footnotesize{Dataset}                                                              & \footnotesize{value}                                                       \\ \hline
\footnotesize{Start time}                & \footnotesize{Yelp}                                                                 & \footnotesize{Check-in}                                                    & \footnotesize{Start time}                                                                        & \footnotesize{Yelp}                                                                 & \footnotesize{Check-in}                                                    \\ \hline
\footnotesize{End time}                  & \footnotesize{Uniform distribution}                                                               & \footnotesize{Uniform distribution}                                                      & \footnotesize{End time}                                                                     & \begin{tabular}[c]{@{}c@{}}\footnotesize{Uniform distribution}\end{tabular}              & \begin{tabular}[c]{@{}c@{}}\footnotesize{Uniform distribution}\end{tabular}     \\ \hline
\footnotesize{Energy capacity}           & \begin{tabular}[c]{@{}c@{}}\footnotesize{Renewable energy sharing}\end{tabular} & \begin{tabular}[c]{@{}c@{}}\footnotesize{Provided energy}\end{tabular} & \footnotesize{Energy capacity}                                                                   & \begin{tabular}[c]{@{}c@{}}\footnotesize{Renewable energy sharing}\end{tabular} & \begin{tabular}[c]{@{}c@{}}\footnotesize{Consumed  energy}\end{tabular} \\ 
\hline
% \footnotesize{Reliability score}         & \footnotesize{Carat}                                                                & \footnotesize{Entropy}                                                      & \footnotesize{Hard deadline}                                                                     & \begin{tabular}[c]{@{}c@{}}\footnotesize{Monte carlo}\end{tabular}              & \begin{tabular}[c]{@{}c@{}}\footnotesize{Monte carlo}\end{tabular}     \\ \hline

\end{tabular}
\caption{Parameters of the experiments setting}
%\vspace{-.6cm}
\label{tab:simparam}
\end{table*}

\begin{figure*}[t!]
    \centering
    \subfloat[]{\includegraphics[width=.26\textwidth]{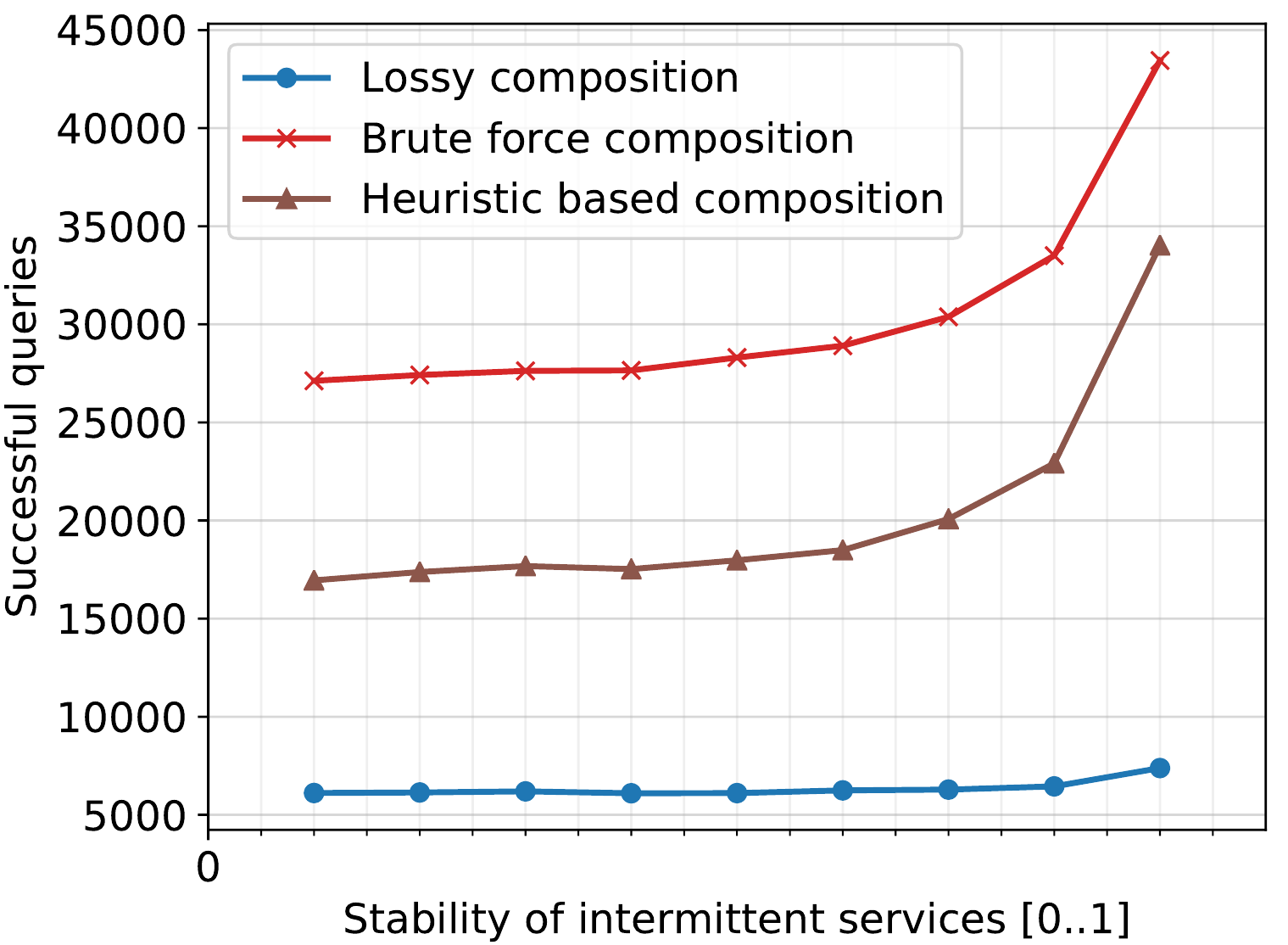}}
    \hfill
    \subfloat[]{\includegraphics[width=.26\textwidth]{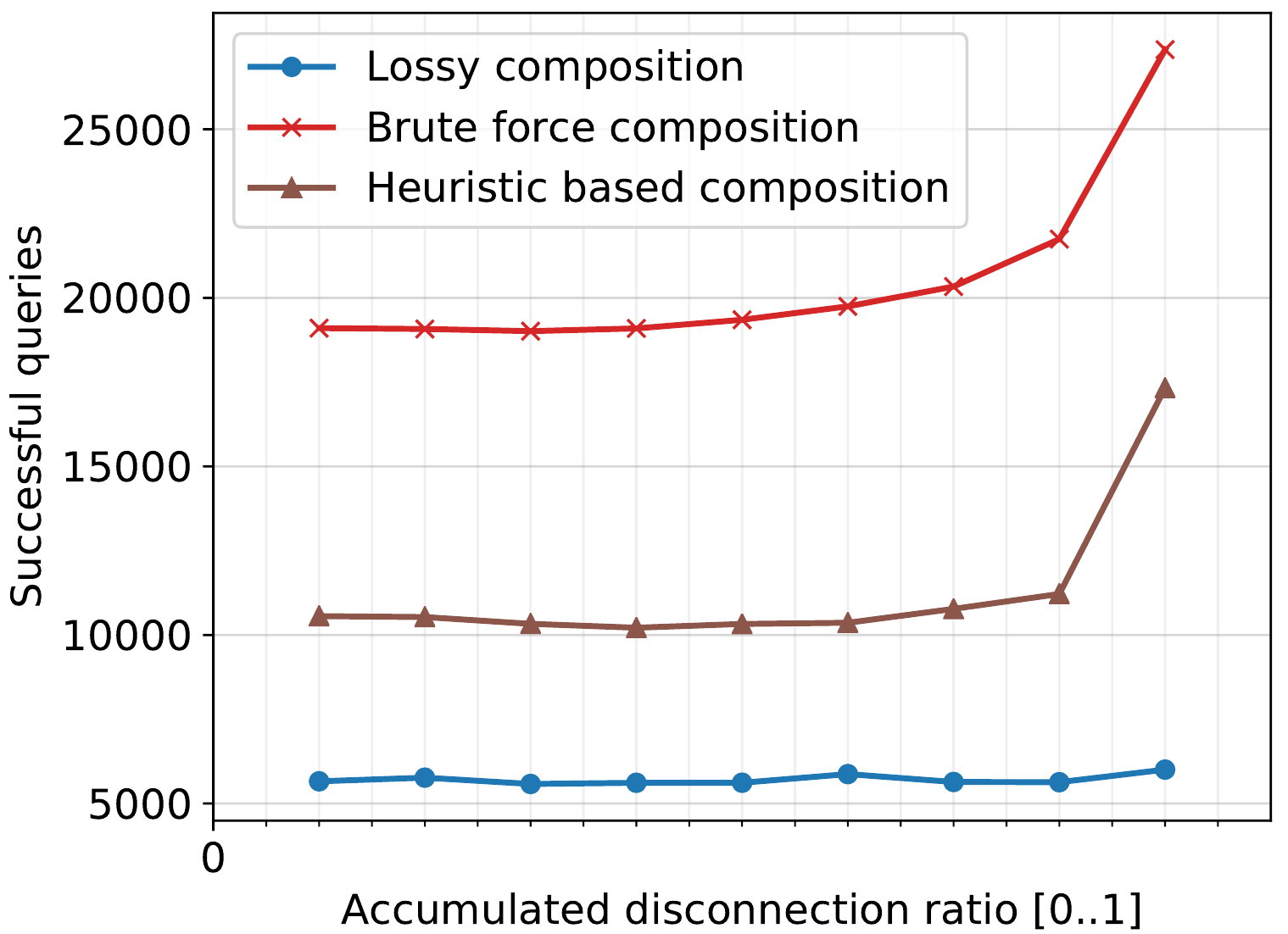}}
    \hfill
    \subfloat[]{\includegraphics[width=.26\textwidth]{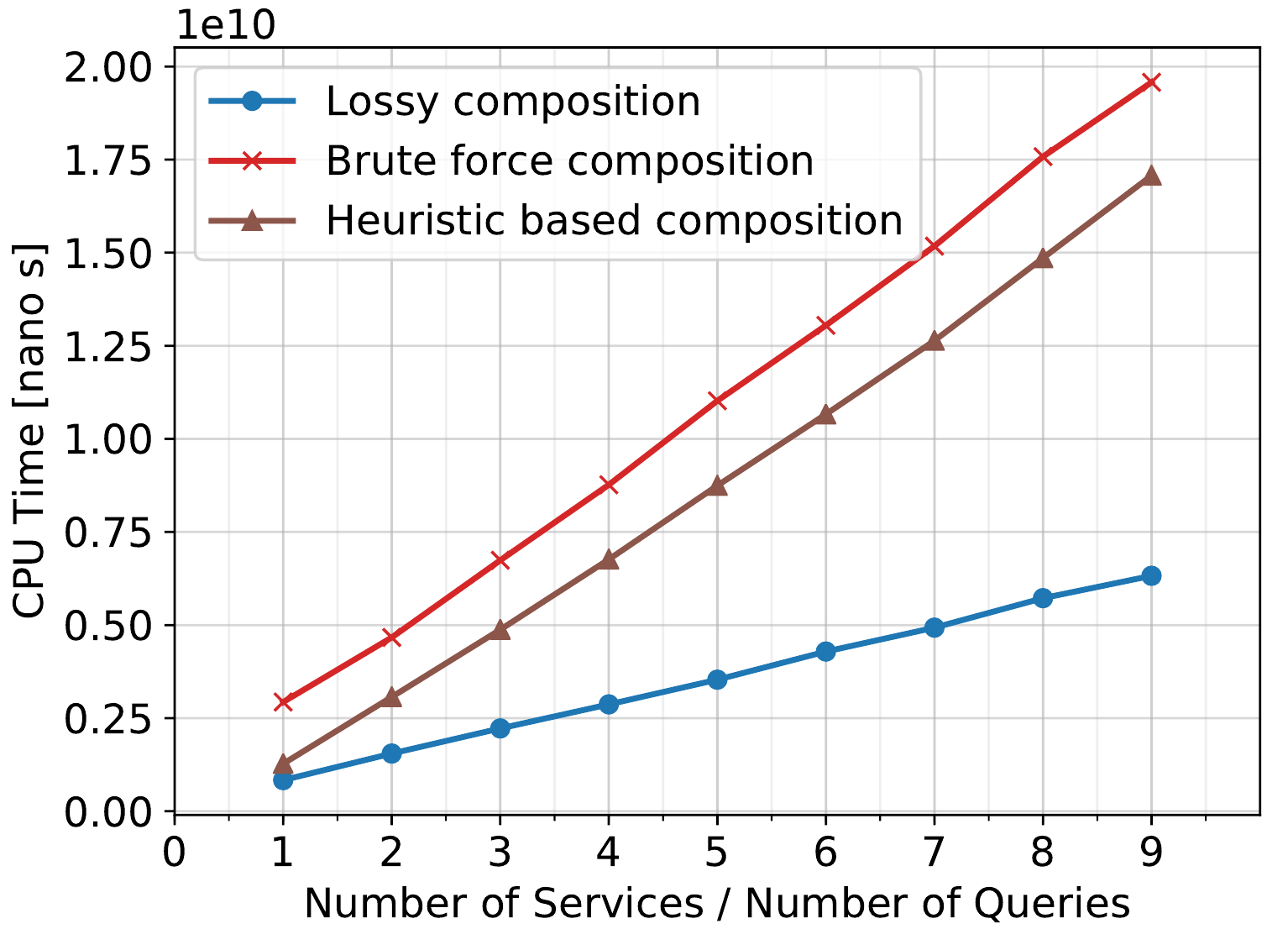}}
    %  %%%%\vspace{-0.35 cm}
        %\vspace{-0.1cm}
    \caption{ \small The effectiveness of fluid composition for short services (a) Successfully served queries Vs stability of intermittent services (b) Successfully served queries Vs length of intermittent disconnections
    (c) CPU time for short services}
         %\vspace{-0.4cm}
    \label{fig:effectivenessShort}
% %%%%\vspace{-0.6 cm}
\end{figure*}

\begin{figure*}[t!]
    \centering
    \subfloat[]{\includegraphics[width=.26\textwidth]{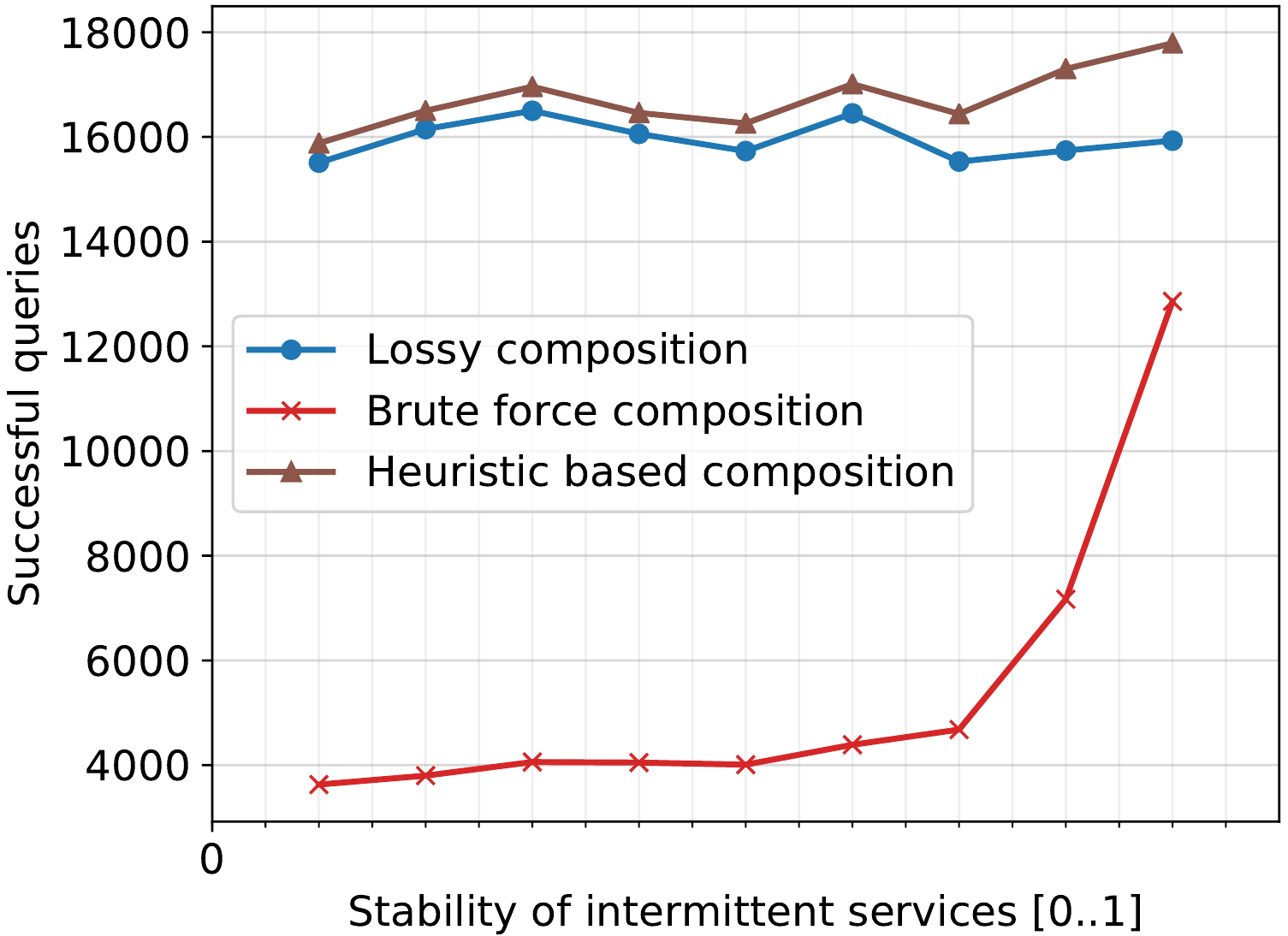}}
    \hfill
    \subfloat[]{\includegraphics[width=.26\textwidth]{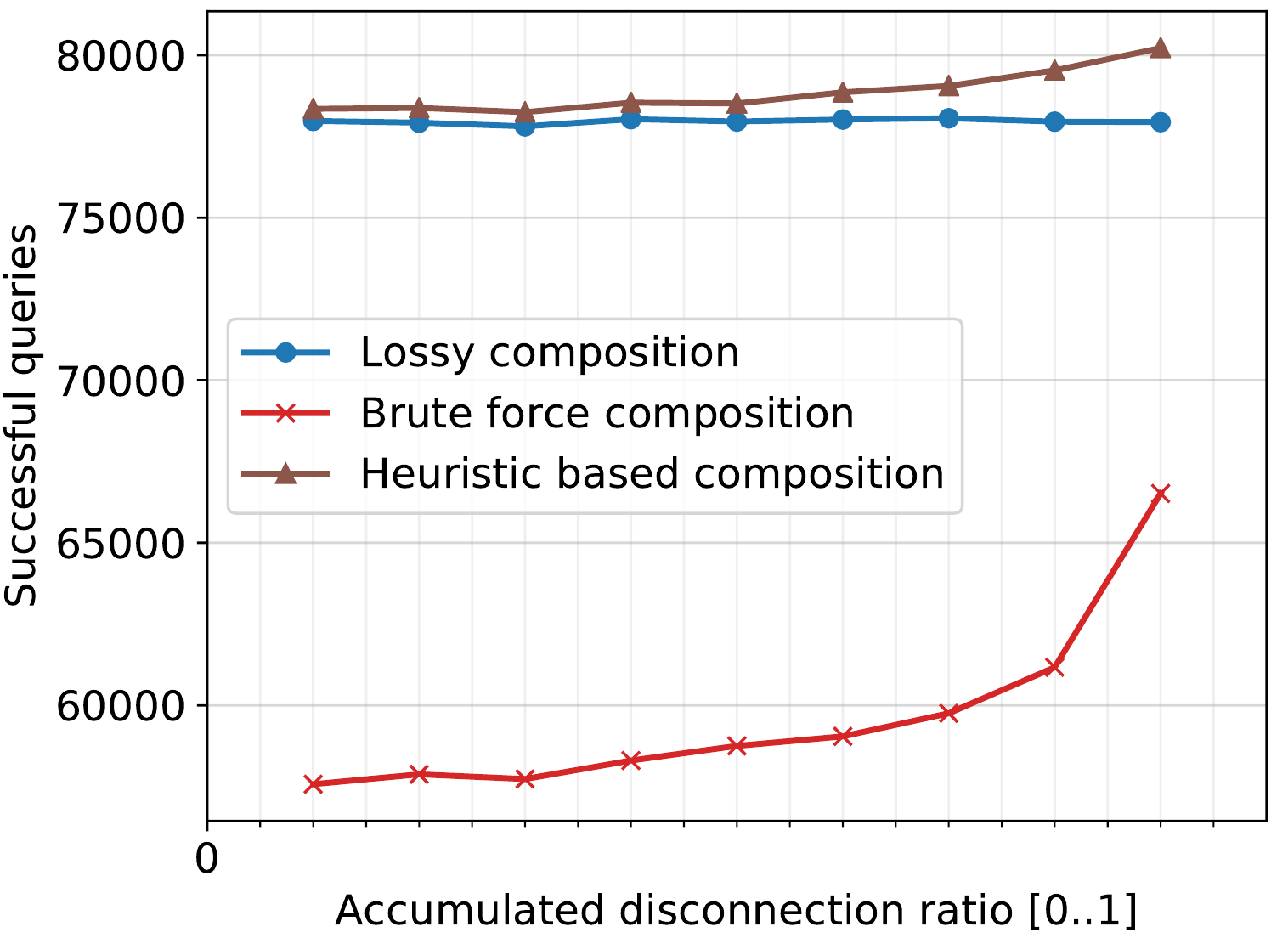}}
    \hfill
    \subfloat[]{\includegraphics[width=.26\textwidth]{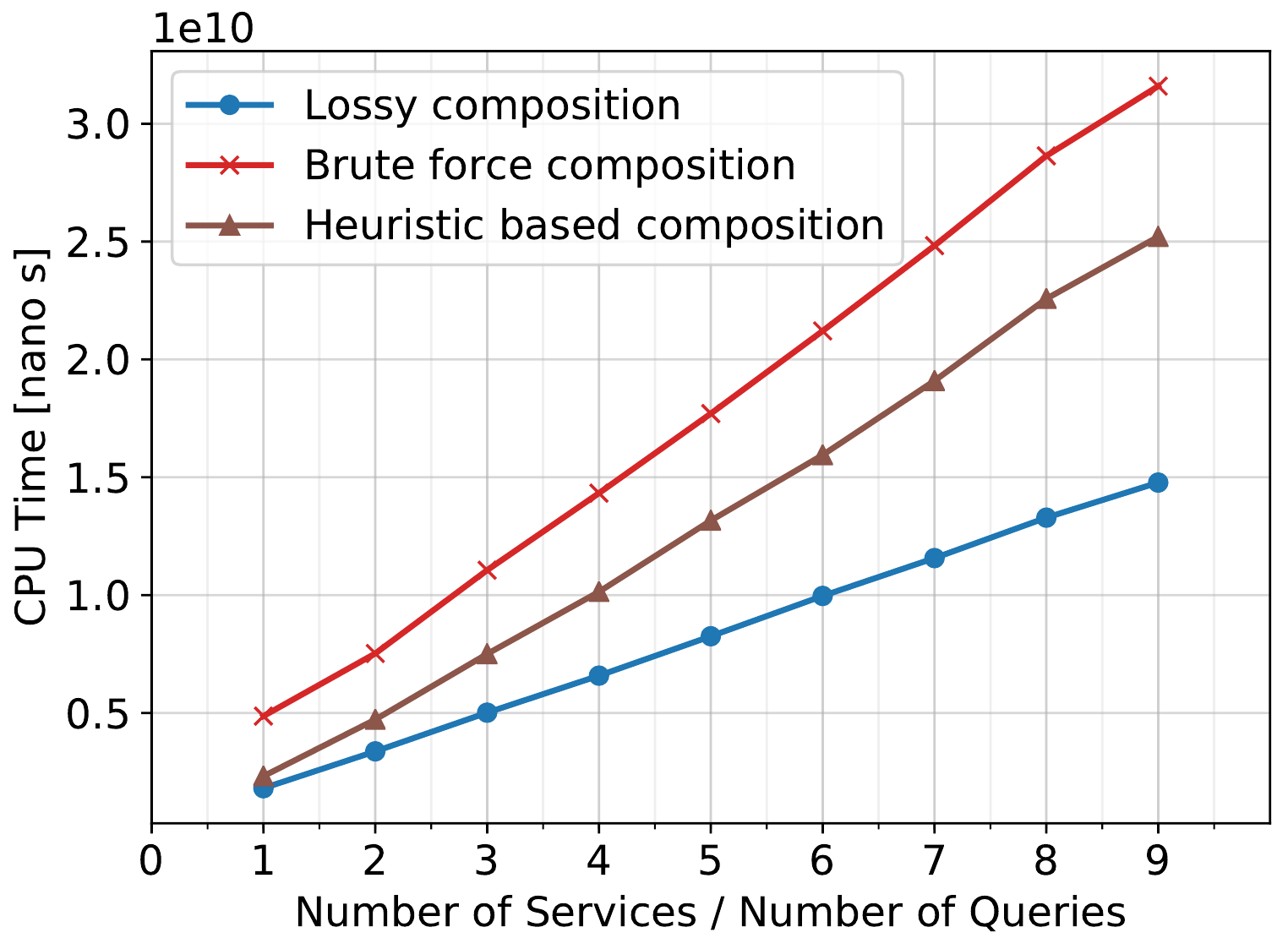}}
    %  %%%%\vspace{-0.35 cm}
        \vspace{-0.1cm}
    \caption{ \small The effectiveness of fluid composition for long services (a) Successfully served queries Vs stability of intermittent services (b) Successfully served queries Vs length of intermittent disconnections
    (c) CPU time for long services}
        \vspace{-0.4cm}
    \label{fig:effectivenessLong}
% %%%%\vspace{-0.6 cm}
\end{figure*}
We propose a heuristic-based composition algorithm using the temporal knapsack algorithm. The objective of the heuristic is to find the optimal composition of intermittent energy services solving the trade-off between composition accuracy and runtime efficiency. Algorithm \ref{alg:STcompo}  presents the pseudocode of the heuristic-based fluid composition. Given a set of composable services based on their advertisement $NearbyS$, the heuristic verifies the mobility patterns {\em (i.e., availability $A$ and provision intermittence $In$ in section \ref{sysmdlpb})} for all composable services. The heuristic does not consider short disconnections (see  {\em short disconnections} in figure \ref{fig:heuristic}). If the aforementioned stability metric $STB$ is higher than a predefined stability threshold $\mu$ the candidate service will be discarded. However, if the candidate service has a stable mobility pattern with few long disconnections according to the accumulated disconnection score $ADis$, the heuristic  then finds one or more substitutes for each long disconnection (Lines 1-8) (see  {\em long disconnections} in figure \ref{fig:heuristic}). \textit{The heuristic algorithm establishes a connection with the intermittent service and its substitute services proactively}. This composition strategy avoids the addition of more chunks by considering the service and its substitutes as a single service (Line 9). The heuristic chunks the request duration based on the advertised time intervals of available services (Lines 10-16). The initial services and their substitutes are temporally composed using the initial chunks and the 0/1 knapsack algorithm (Lines 17-25). Relying on the initial chunking and toleration of short disconnections makes the heuristic-based composition runtime efficient compared to the brute-force which defines fine grained chunking considering all disconnections (see section \ref{sec:scalevi}). The substitute services provide a patch to recover all the disconnections of the initial advertised service. This lightweight process to reconstruct the initial advertised service is the main reason of increasing the accuracy of the heuristic based composition.    

% \begin{figure}
%     \centering
%     \includegraphics[width=.29\textwidth]{Figures/to_graph.pdf}
%     \vspace{-0.2cm}
%     \caption{\small The effectiveness of static and fluid compositions of crowdsourced energy services}
%     \vspace{-0.6cm}
%     \label{fig:statvsfluid}
% \end{figure}
% \vspace{-0.5cm}
\vspace{-0.22cm}
\section{Experiments}
\vspace{-0.12cm}
We evaluate the \textit{effectiveness} of the proposed fluid composition approach by assessing the composition performance in a \textit{failure-prone} crowdsourced mobile  IoT environment. We also evaluate the \textit{scalability} of the fluid composition algorithm by measuring the computation time while varying the requests number. We compare the proposed approach with two state of the art composition algorithms, (i.e., static spatio-temporal composition \cite{lakhdari2020composing} and a lossy  Web service composition \cite{medjahed2003composing}) and with the brute-force approach.
\vspace{-0.22cm}
\subsection{Datasets and experiment environment}
\vspace{-0.12cm}
% All algorithms are implemented in Java. The experiments are conducted on a 3.60 GHz Core i7 processor and 8 GB RAM under Windows 10.
% 
\textit{To the best of our knowledge, it is challenging to find dataset about the energy wireless transfer among human-centric IoT devices}. We create a crowdsourced IoT environment scenario close to reality. We mimic the energy harvesting and sharing behavior of the crowd by utilizing \textit{QLD smartgrid}\footnote{https://data.gov.au/dataset}, an energy sharing smart-grid of 25 houses in Queensland Australia equipped with solar panels. These houses harvest energy from the solar panels, consume energy and share their spare energy by pushing it back to the smartgrid ({\em i.e., energy providers}) to cater for other houses if their produced energy is not sufficient for their requirements. 
%They are considered as energy providers. 
Similarly, the energy requirements of a request $Q.RE$ are also generated from the daily energy consumption of the houses. \textit{QLD smartgrid} contains the daily energy data of the 25 houses in Queensland for two years [April 2012 to March 2014]. Energy consumption and production is recorded every 30 minutes. Each house has 730 records. Each record has $48$x$2$ fields for the produced and the consumed energy at every 30 minutes. In our experiment, we define  the energy service QoS parameters, the deliverable energy capacity $DEC$ and the intensity of the transferred current $I$ from \textit{QLD smartgrid} dataset. The  wireless transmission success rate $Tsr$ QoS parameter is  randomly generated.

% We use this dataset to mimic energy harvesting and sharing behavior. A set of houses harvest energy from their solar panels,

We use {\em Yelp}\footnote{https://www.yelp.com/dataset} dataset to simulate the spatio-temporal features of crowdsourced energy services and requests. This dataset contains several information about the crowd's behavior in different venues in multiple metropolitan cities e.g., coffee shops, restaurants, libraries, etc. People may check-in, rate and recommend these venues. In our experiment, we only focus on people's check-ins information into confined areas e.g., coffee shops. We consider the $Yelp\_checkin$ table. For each venue ($business\_id$), we extract the crowd size ($Checkins$) at each hour ($hour$) of the day ($weekday$). We assume these people as IoT users. They may offer energy services from their wearables while staying in a confined area. We define spatio-temporal features of energy services by  generating the check-in and check-out timestamps of customers to confined areas using the  previously extracted data from  $Yelp\_checkin$ table. For example, the start time $st$ of an energy service from an IoT user is the time of their check-ins into a coffee shop. Energy request time $Q.t$ and duration $Q.du$ are also generated from check-in and check-out times of customers. 

We match the two datasets \textit{Yelp} and \textit{QLD smartgrid} by randomly mapping each energy service $S_i$ starting at $st_i$ and ending at $et_i$ with the produced energy during the same period from one of the $25$x$730$ records. Similarly, the required energy $Q.RE$ is also randomly selected from one of the $25$x$730$ records according to the request duration $Q.du$. 
We normalize all the energy measurement values for all records from {\em Watt hour to miliampere hour (mAh)} to mimic the energy provided and consumed by IoT devices e.g., smartphone and wearables. Table \ref{tab:simparam} recapitulates the experiments parameters. To mimic the intermittent behavior of wireless energy delivery in confined areas, we augment our dataset by generating random disconnections for all the services. We implement a parameterized randomizer for all the energy services. The disconnections are monitored using two parameters, their {\em frequencies} and their {\em lengths}.
% \cite{kar2009distributed}
\vspace{-0.22cm}
\subsection{Effectiveness}
\vspace{-0.12cm}
\begin{figure}
    \centering
    \includegraphics[width=.29\textwidth]{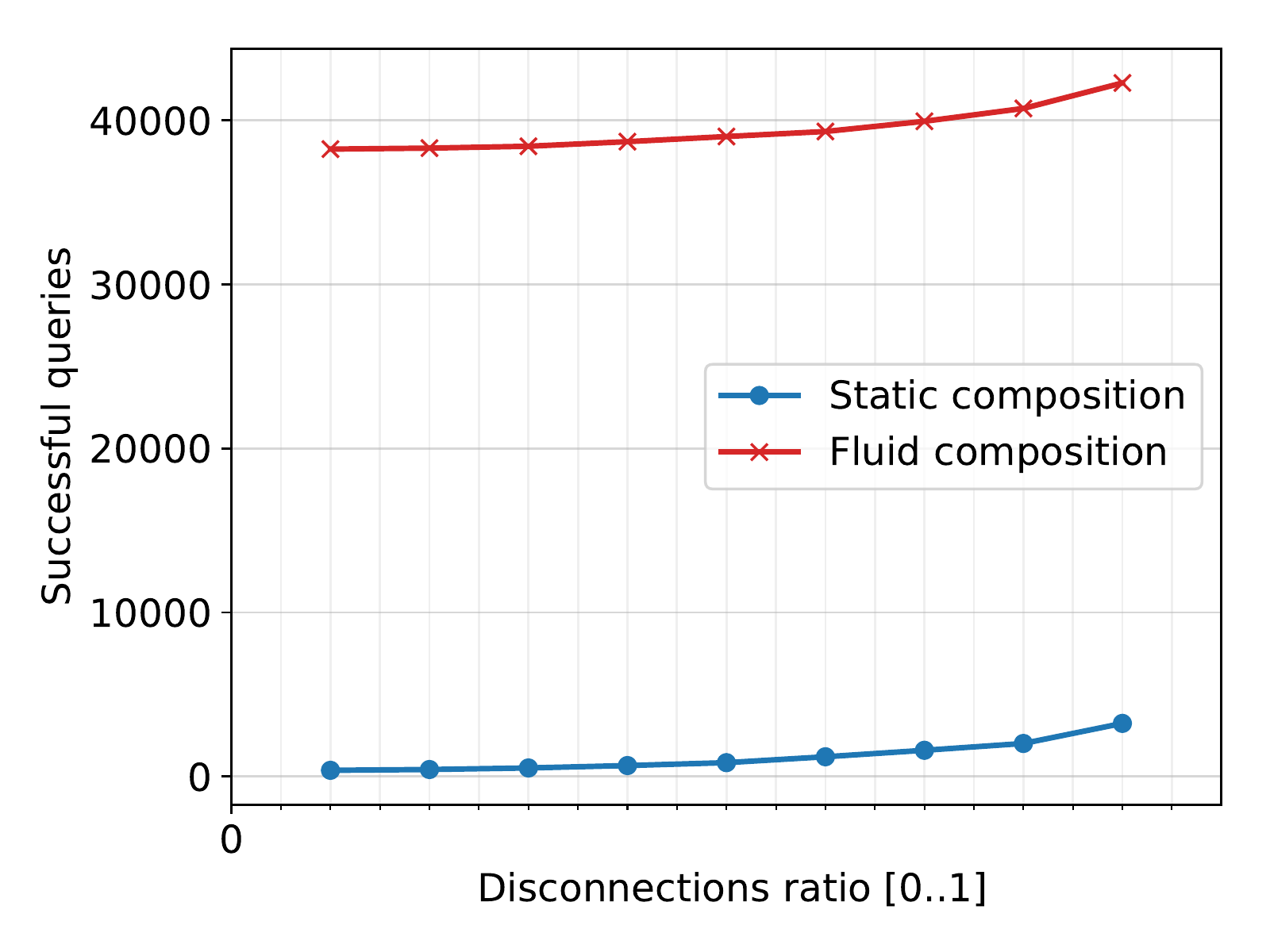}
    \vspace{-0.2cm}
    \caption{\small The effectiveness of static and fluid compositions of crowdsourced energy services}
    \vspace{-0.6cm}
    \label{fig:statvsfluid}
\end{figure}
We investigate the effectiveness of the fluid composition by comparing the number of successfully served requests by each algorithm. We use two features of intermittent services to evaluate the performance. The stability of an intermittent service which reflects the number of disconnections. The accumulated disconnection ratio which represents the intermittence and the lost energy. We compare the heuristic-based fluid composition with the static composition \cite{lakhdari2020composing} (see Figure \ref{fig:statvsfluid}). The goal of this initial comparison is to highlight the non usability of static spatio-temporal composition algorithms in a highly dynamic crowdsourced IoT environment. We run the static and fluid composition algorithms on energy services and requests with different duration length, provided, and requested energy. We increase the dynamicity by varying the number of services' disconnections gradually. It is obvious that the static composition performs poorly with intermittent services. This behavior is expected because the algorithm has been designed for static energy services. All intermittent services are not selected which is reflected on the poor performance of the static composition even when the dynamicity of the crowdsourced IoT environment is low.

Second, we run the brute force, the heuristic-based, and the traditional composition algorithm \cite{medjahed2003composing} on short and long services to understand the performance behavior in more details. In short services, the more stable services, the better performance of all algorithms (see Figure \ref{fig:effectivenessShort} (a)). 
The brute force composition has the best performance because it considers all the possible chunks.
%(see Figure \ref{fig:Chunking} in section \ref{ftrs}). 
The traditional (i.e., lossy) composition has the least performance regardless the stability of the service due to {\em filtering out the highly intermittent services} from the beginning. The heuristic performance is comparable with the brute force because {\em it does not filter out highly intermittent services and attempts to find substitutes at each disconnection}. Figure \ref{fig:effectivenessShort} (b) also presents the performance behavior of the composition algorithms against the accumulated disconnection time of intermittent services. The brute force and the heuristic perform well, and they fulfill the requirements of a large number of requests unlike the lossy conmposition. A considerable accumulated disconnection time means the creation of several sub services which can be discovered by the brute force and the heuristic.%-based composition. 

Surprisingly, the brute force has a poor performance against long intermittent services ( see Figure \ref{fig:effectivenessLong} (a) and (b)). Unlike short services, disconnected long services have the same wireless current intensity. If two long services are not composable, their sub services also cannot be composable. Thus, even the exhaustive chunking, 0/1 knapsack cannot find composable services at these chunks. The heuristic and the lossy algorithm consider longer chunks because their chunking method relies only on the initial advertisement of services.  They also assess intermittent services to select only long stable services to find stable substitutes.    

\vspace{-0.22cm}
\subsection{Scalability}\label{sec:scalevi}
\vspace{-0.12cm}
We evaluate the scalability of the fluid composition to ensure an efficient deployment of the proposed framework on an edge-based IoT architecture. The edge-based IoT coordinator 
% will be running at the edge (e.g., a router in a coffee shop as previously mentioned in the motivating scenario) 
composes intermittent energy services in a dynamic crowdsourced IoT environment. We compare the average execution time of three different composition techniques. We run each algorithm on 480000 crowdsourced energy services. We vary the number of available services for each request from 1 to 9. figures \ref{fig:effectivenessShort} (c) and \ref{fig:effectivenessLong} (c) represent the behavior of the three different algorithms for short and long services. The results show that the execution time increases as the $(number~of~services/number ~of ~queries)$ increases. This performance behavior is expected from all the composition techniques because of the increase in the number of services from IoT users.

The brute force composition takes longer execution time for short and long services. The lossy composition does not require a long execution time because it filters out highly intermittent services before starting the composition. However, the brute force considers every disconnection and defines new sub-services which generate multiple chunks 
%(see Figure \ref{fig:Chunking} in section \ref{ftrs})
. In addition to the time required by the 0/1 knapsack algorithm at each chunk. The heuristic-based composition takes longer execution time than the lossy composition because it does not filter out the intermittent services. The heuristic algorithm attempts to patch the disconnections of intermittent services by exploring the nearby available services which take considerable time before the composition. However, the heuristic-based composition relies on the initial chunking based on the advertisement of available services. Applying the 0/1 knapsack algorithm on the initial chunks takes less time than applying the algorithm on all the new chunks generated from all disconnections.

\vspace{-0.22cm}
\section{Related work}
\vspace{-0.12cm}
% \textcolor{red}{Give related work about wireless energy transfer, service composition in IoTenvironment, composition in dynamic environment and existing energy composition in deterministic environment. Then conculde they are not applicable for the unique feaure of energy services, unique requirement for energy composibility and the intermittent behavior}
Energy consists of a significant challenge in many wireless application domains, including IoT and wireless sensor networks. Self-harvesting energy from natural sources using wearables such as body movement and heat provides a significant source of energy~\cite{seneviratne2017}. A new emerging body of research attempts to integrate harvesting energy into designing IoT objects~\cite{gorlatova2014movers}\cite{khalifa2017harke}. {\em Energous Wattup} provides a prototype for wireless charging by eliminating direction requirement which creates an opportunity of sharing energy between IoT devices. Service computing is a key enabler for wireless energy sharing. Several service composition techniques have been proposed in pervasive and ubiquitous computing recently. such as cloud computing, sensor-cloud services~\cite{neiat2017crowdsourced} and social networks~\cite{bouguettaya2017service}. In sensor-cloud, services are composed according to their spatio-temporal features. They also must fulfill consumer preferences (QoS). Neiat et al.~~\cite{neiat2017crowdsourced} design and implement a spatio-temporal service composition framework. The spatio-temporal service composition framework has been extended to describe and compose region services like WiFi hotspots. Their goal is to provide the most convenient trip plan from point A to point B with the best crowdsourced WiFi coverage.  User preferences are used to define the spatial and temporal composability models of segment services \cite{lakhdari2020composing}. 
%~\cite{wang2014constraint}
Existing composition techniques are not applicable for intermittent crowdsourced energy services due to the unique features of energy services, unique requirement for energy composability and the intermittent behavior. The proposed approach adds a new contribution to the existing related studies \cite{Previouswork11}\cite{lakhdari2020composing}. The proposed service model captures functional and QoS requirements of energy services. The proposed spatio-temporal composition framework considers the intermittent behavior of crowdsourced energy services in a dynamic IoT environment. The proposed heuristic algorithm estimates the impact of services' intermittence to select the optimal composition which delivers the required energy within the shortest disconnetion time between the consuming device and the providers. This work is one of the earliest contributions to compose mobile crowdsourced energy services in a dynamic IoT environment.% in the field of service computing. 
\vspace{-0.22cm}
\section{Conclusion}
\vspace{-0.12cm}
We propose a novel fluid spatio-temporal composition framework to crowdsource energy services from IoT devices. %The objective is to meet users' energy requirements by composing intermittent crowdsourced energy services in a dynamic environment. 
We develop a composition approach which estimates the stability of spatio-temporal composition plans and provides the optimal composition in the predefined time interval and location. We conduct a set of experiments to investigate the scalability and the effectiveness of the proposed composition technique. %In an IoT environment, the energy services might fail due to the providing devices usage. 
Results show that the proposed approach is able to provide the most stable composition with the minimal disconnection time. In future work, we will develop a semi reactive composition of mobile  energy services.% i.e., dynamic composition in confined places.
\vspace{-0.22 cm}

% \section*{Acknowledgment}

\bibliographystyle{IEEEtran}
\bibliography{IEEEabrv, ref}

\begin{thebibliography}{10}
\providecommand{\url}[1]{#1}
\csname url@samestyle\endcsname
\providecommand{\newblock}{\relax}
\providecommand{\bibinfo}[2]{#2}
\providecommand{\BIBentrySTDinterwordspacing}{\spaceskip=0pt\relax}
\providecommand{\BIBentryALTinterwordstretchfactor}{4}
\providecommand{\BIBentryALTinterwordspacing}{\spaceskip=\fontdimen2\font plus
\BIBentryALTinterwordstretchfactor\fontdimen3\font minus
  \fontdimen4\font\relax}
\providecommand{\BIBforeignlanguage}[2]{{%
\expandafter\ifx\csname l@#1\endcsname\relax
\typeout{** WARNING: IEEEtran.bst: No hyphenation pattern has been}%
\typeout{** loaded for the language `#1'. Using the pattern for}%
\typeout{** the default language instead.}%
\else
\language=\csname l@#1\endcsname
\fi
#2}}
\providecommand{\BIBdecl}{\relax}
\BIBdecl
\renewcommand{\BIBentryALTinterwordstretchfactor}{4}

\bibitem{perera2014survey}
C.~Perera \emph{et~al.}, ``A survey on internet of things from industrial
  market perspective,'' \emph{IEEE Access}, vol.~2, 2014.

\bibitem{bouguettaya2017service}
A.~Bouguettaya \emph{et~al.}, ``A service computing manifesto: the next 10
  years,'' \emph{Communications of the ACM}, 2017.

\bibitem{ren2015exploiting}
J.~Ren \emph{et~al.}, ``Exploiting mobile crowdsourcing for pervasive cloud
  services: challenges and solutions,'' \emph{IEEE Comm.}, 2015.

\bibitem{ahabak2015femto}
K.~Habak \emph{et~al.}, ``Femto clouds: Leveraging mobile devices to provide
  cloud service at the edge,'' in \emph{IEEE CLOUD}, 2015.

\bibitem{zhang2015incentives}
X.~Zhang \emph{et~al.}, ``Incentives for mobile crowd sensing: A survey,''
  \emph{IEEE Communications Surveys \& Tutorials}, 2015.

\bibitem{dhungana2020peer}
A.~Dhungana and E.~Bulut, ``Peer-to-peer energy sharing in mobile networks:
  Applications, challenges, and open problems,'' \emph{Ad Hoc Networks}, 2020.

\bibitem{bulut2018crowdcharging}
E.~Bulut \emph{et~al.}, ``Is crowdcharging possible?'' in \emph{ICCCN}, 2018.

\bibitem{gorlatova2014movers}
M.~Gorlatova \emph{et~al.}, ``Movers and shakers: Kinetic energy harvesting for
  the internet of things,'' in \emph{ACM SIGMETRICS Performance Evaluation
  Review}, 2014.

\bibitem{abusafia2020incentive}
A.~Abusafia \emph{et~al.}, ``Incentive-based selection and composition of iot
  energy services,'' \emph{arXiv preprint arXiv:2007.09985}, 2020.

\bibitem{Previouswork11}
A.~Lakhdari \emph{et~al.}, ``Crowdsourcing energy as a service,'' in
  \emph{Springer ICSOC}, 2018.

\bibitem{lakhdari2020composing}
A.~Lakhdari \emph{et~al.}, ``Composing energy services in a crowdsourced iot
  environment,'' \emph{IEEE TSC}, 2020.

\bibitem{lemos2016web}
A.L. Lemos \emph{et~al.}, ``Web service composition: a survey of techniques and
  tools,'' \emph{ACM Computing Surveys}, 2016.

\bibitem{waxman2006coffee}
L.~Waxman, ``The coffee shop: Social and physical factors influencing place
  attachment,'' \emph{Jrnl. of Interior Design}, 2006.

\bibitem{medjahed2003composing}
B.~Medjahed \emph{et~al.}, ``Composing web services on the semantic web,''
  \emph{The VLDB journal}, vol.~12, no.~4, pp. 333--351, 2003.

\bibitem{yang2015mobility}
Z.~Yang \emph{et~al.}, ``Mobility increases localizability: A survey on
  wireless indoor localization using inertial sensors,'' \emph{ACM Computing
  Surveys}, 2015.

\bibitem{khalifa2017harke}
S.~Khalifa \emph{et~al.}, ``Harke: Human activity recognition from kinetic
  energy harvesting data in wearable devices,'' \emph{IEEE Transactions on
  Mobile Computing}, 2017.

\bibitem{na2018energy}
W.~Na \emph{et~al.}, ``Energy-efficient mobile charging for wireless power
  transfer in iot networks,'' \emph{IEEE the IoT Journal}, 2018.

\bibitem{gonzalez2008understanding}
M.C. Gonzalez \emph{et~al.}, ``Understanding individual human mobility
  patterns,'' \emph{nature}, 2008.

\bibitem{deng2016constraints}
S.~Deng \emph{et~al.}, ``Constraints-driven service composition in mobile cloud
  computing,'' in \emph{IEEE ICWS}, 2016.

\bibitem{bartlett2005temporal}
M.~Bartlett \emph{et~al.}, ``The temporal knapsack problem and its solution,''
  in \emph{Springer CPAIOR}, 2005.

\bibitem{seneviratne2017}
S.~Seneviratne \emph{et~al.}, ``A survey of wearable devices and challenges,''
  \emph{IEEE Communications Surveys \& Tutorials}, 2017.

\bibitem{neiat2017crowdsourced}
A.G. Neiat \emph{et~al.}, ``Crowdsourced coverage as a service: Two-level
  composition of sensor cloud services,'' \emph{IEEE TKDE}, 2017.

\end{thebibliography}
\vspace{-0.25 cm}

\end{document}